\documentclass[review, 3p]{elsarticle} 

\usepackage{lineno,hyperref}
\modulolinenumbers[5]

\journal{Physica A}

\usepackage{graphicx}  
\graphicspath{{./Figure/}}        
\usepackage{multirow}             
\usepackage{color}                
\usepackage{amsmath}              
\usepackage{amssymb}              
\usepackage{lscape}               

\begin{document}

\begin{frontmatter}

\title{Exclusive queueing model including the choice of service windows} 


\author[mymainaddress]{Masahiro Tanaka\corref{mycorrespondingauthor}}
\cortext[mycorrespondingauthor]{Corresponding author}
\ead{masatanaka@g.ecc.u-tokyo.ac.jp}

\author[mysecondaryaddress,mymainaddress]{Daichi Yanagisawa}
\ead{tDaichi@mail.ecc.u-tokyo.ac.jp}

\author[mysecondaryaddress,mymainaddress]{Katsuhiro Nishinari}
\ead{tknishi@mail.ecc.u-tokyo.ac.jp}

\address[mymainaddress]{Department of Aeronautics and Astronautics, School of Engineering, The University of Tokyo, 7-3-1, Hongo, Bunkyo-ku, Tokyo, 113-8656, Japan}
\address[mysecondaryaddress]{Research Center for Advanced Science and Technology, The University of Tokyo, 4-6-1, Komaba, Meguro-ku, Tokyo 153-8904, Japan}

\begin{abstract}
In a queueing system involving multiple service windows, choice behavior is a significant concern. 
This paper incorporates the choice of service windows into a queueing model with a floor represented by discrete cells. 
We contrived a logit-based choice algorithm for agents considering the numbers of agents and the distances to all service windows. 
Simulations were conducted with various parameters of agent choice preference for these two elements and for different floor configurations, including the floor length and the number of service windows. 
We investigated the model from the viewpoint of transit times and entrance block rates. 
The influences of the parameters on these factors were surveyed in detail and we determined that there are optimum floor lengths that minimize the transit times. 
In addition, we observed that the transit times were determined almost entirely by the entrance block rates. 
The results of the presented model are relevant to understanding queueing systems including the choice of service windows and can be employed to optimize facility design and floor management. 
\end{abstract}


\end{frontmatter}


\section{Introduction}
\label{sec:Introduction}
There are a considerable number of pedestrian queueing systems, including cash registers in shops, automated teller machines in banks, and ticket-vending machines in stations. 
It is important to investigate queueing systems for several reasons. 
One reason is that pedestrians in queues usually feel discontented to be doing nothing but waiting. 
For example, many people who want to enjoy popular attractions in an amusement park have to wait in queues. 
Another reason is that queues have unfavorable effects on the surrounding environment. 
For example, when an attractive device is on sale, enthusiasts form queues around shops and obstruct roads.

One major strategy to deal with pedestrian queueing systems is queueing theory \cite{Blanchard1990}. 
Queueing theory originates from the work of Erlang \cite{Erlang1909}, who explored telephone exchange systems. 
Numerous advances have subsequently been made, such as Kendall's notation \cite{Kendall1953}, Burke's theorem \cite{Burke1956}, the Jackson network \cite{Jackson1957}, and Little's theorem \cite{Little1961}. 
Recently, queueing theory has been applied to pedestrian and vehicle traffic \cite{Helbing2005,Helbing2006}, the Internet \cite{Erramilli1996,Karol1987}, and logistics \cite{Taniguchi1999}.

Another sophisticated method to analyze pedestrian queueing systems is the totally asymmetric simple exclusion process (TASEP).
TASEP is an elementary mathematical model established by MacDonald and Gibbs \cite{MacDonald1968,MacDonald1969} to examine the kinetics of biopolymers and polypeptides.
Its characteristics have been explored \cite{Blythe2007,Schadschneider2013,Woelki2008}, and it has been applied in many fields, \textit{e.g.}, car traffic models \cite{Schadschneider2013,Nagel1992,Schreckenberg1995,Schadschneider1997} and pedestrian queueing models \cite{Helbing1991}.

Recently, the exclusive queueing model \cite{Yanagisawa2010,Yanagisawa2013,Arita2014,Tanaka2016} was developed by combining TASEP with a normal queueing model. 
This model reproduces pedestrian queueing systems better than conventional normal queueing models. 
Even though the spatial factor is considered in the exclusive queueing model, the choice of service windows has not been incorporated. 
Yanagisawa et al. \cite{Yanagisawa2013} created a model with exclusive queueing passages to service windows and normal queueing waiting lanes prior to the distribution of pedestrians to service windows. 
Their model distributes pedestrians to the service windows if they are vacant.
However, there are many situations where pedestrians need to choose which service windows they should approach. 
Therefore, in this study, we created an exclusive queueing model with a choice of service windows, involving one entrance and multiple service windows. 
Its simplicity accommodates a vast array of applications, such as security-check areas in airports and ticket gates in amusement parks. 
Therefore, this study can make a substantial contribution to the management of systems similar to those in our model.

There are two types of choices in our daily lives: choices given by a managing authority and choices determined by the pedestrians themselves. 
One example of the former is the distribution of users in a call center \cite{Armony2010,Doroudi2011,Li2016}. 
A stair--escalator choice provides an example of the latter \cite{Ji2013}. 
There are multiple situations where a managing staff helps an pedestrian choose; however, it is important to investigate how pedestrians choose their target service windows. 
Therefore, we consider a combined situation, that is, we adopted a system in which the pedestrians' choices were self-determined with staff support and addressed the conditions that attained efficient management. 

The choice of exit for pedestrians in evacuation literature has been investigated by many researchers. 
When pedestrians choose exits, there are several factors that influence how they proceed. 
According to questionnaires and experiments, pedestrians tend to emphasize four criteria: (1) the distances to exits, (2) the pedestrian densities around the exits, (3) whether to follow other pedestrians, and (4) the flow of a particular exit, such as an emergency exit or a main entrance \cite{Augustijn-Beckers2010}. 
Multiple simulations have assumed that the distance is the primary factor that pedestrians consider when choosing exits \cite{Ji2013,Augustijn-Beckers2010,Lo2006,Fu2014,Duives2012,Haghani2014,Lovreglio2016}. 
Congestion has also been integrated into these models with various definitions, such as the number of pedestrians or the density of pedestrians around exits \cite{Ji2013,Augustijn-Beckers2010,Lo2006,Fu2014,Haghani2014,Lovreglio2016} and the pedestrian density around other pedestrians \cite{Duives2012,Lovreglio2016}. 
The abovementioned analyses, except for the stair--escalator choice studied by Ji et al. \cite{Ji2013}, were based on evacuation cases; however, there are few studies concerning the management of a queueing system with a choice of service windows. 
Furthermore, even though Blanc explored the probability that pedestrians lined up in shorter queues would suffer longer waiting times \cite{Blanc2009}, to our knowledge, choices affected by queues to target service windows have not been sufficiently studied. 
Therefore, we tackled these two lapses by addressing the floor management of a queueing system and considering the queue to be an important element when choosing a service window. 
Thus, our research provides an opportunity to achieve administrative improvement in a queueing system under various scenarios that are not limited to evacuation cases. 
Our exclusive queueing model including service window choice concentrated on criteria (1) and (2), which are more fundamental and important elements, as a first step, leaving criteria (3) and (4) to future studies.

When there are several factors influencing a discrete choice, a useful method is the logit model, which was proposed by McFadden \cite{McFadden1973}. 
In studies of route choice, the logit model has frequently been used \cite{Ji2013,Duives2012,Haghani2014,Lovreglio2016,Dijkstra2005}. 
We introduced a logit-based choice model for multiple service windows on a floor represented by discrete cells. 
In our model, the number of agents\footnote{We call the pedestrians in our model not ``pedestrians'' but ``agents.''} and the distances to the service windows were used as the choice elements by the agents because these were considered strongly by real pedestrians. 
Thus, the floor management of a queueing system could be addressed. 

Here, a logit-based choice model for multiple service windows on a floor represented by discrete cells is presented. 
The floor model is extended from the partially exclusive queueing model conceived by Yanagisawa et al. \cite{Yanagisawa2013} to a completely exclusive queueing model. 
Further, we added service window choice by agents as a logit-based probability, where agents consider the congestion degrees of all service windows and the distances when they choose a service window. 
This study primarily aims to explore the characteristics of the floor applicable to a variety of situations from the viewpoint of transit times and entrance block rates. 
The results of this article are useful for administrative improvements and to design floor layouts where pedestrians walk and receive services. 

The remainder of this paper is organized as follows. 
The model is described in Sec. 2. 
The floor configuration and the probability of service window choice are articulated. 
The characteristics of our model derived through our simulation are explained in detail in Sec. 3. 
In addition, the effects of various parameters on the performance measures in our model are investigated. 
We present the conclusions and offer suggestions for future studies in Sec. 4. 
\section{Model}
\label{sec:Model}
\subsection{Floor configuration and update rule}
\label{sec:Model | Floor configuration and update rule}
\begin{figure}[tbh]
	\centerline{\includegraphics[width=11cm]{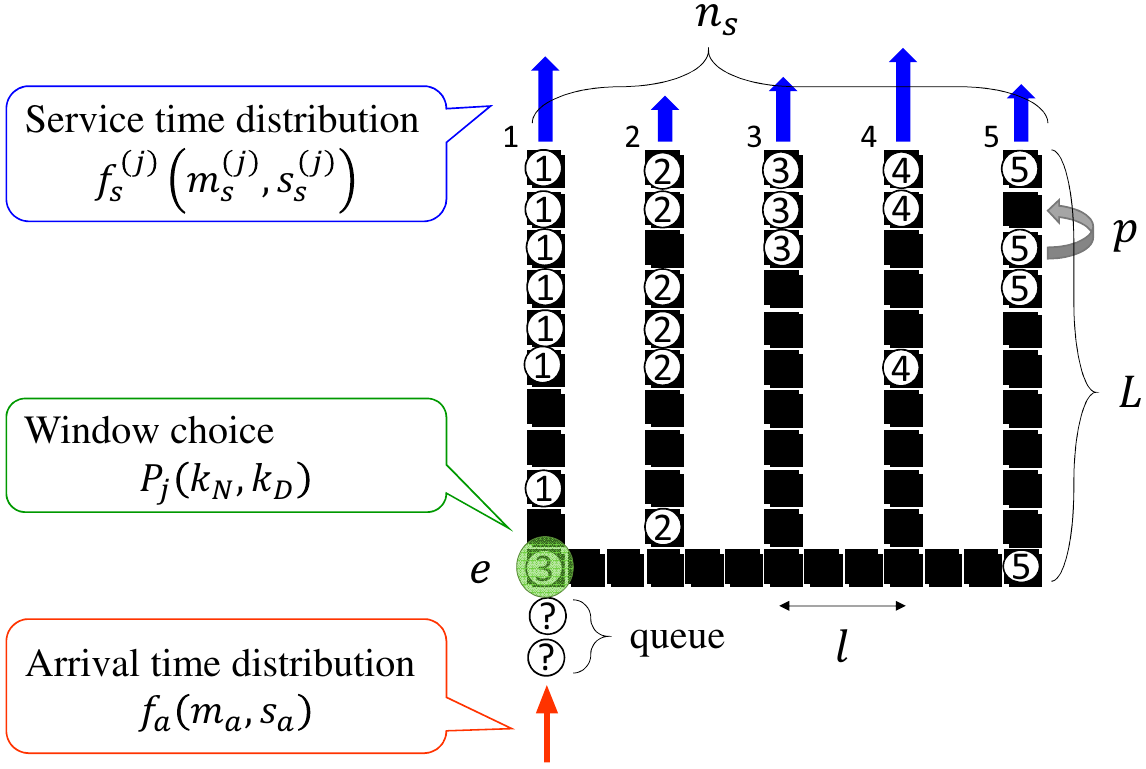}}
	\begin{center}
	\caption{Example state of our simulation for the number of service windows $n_s=5$, the entrance location $e=1$, the window intervals $l=3$, and the floor length $L=11$. The circles indicate agents and the numbers in the circles represent the target service windows of the agents.}
	\label{fig:general floor model}
	\end{center}
\end{figure}

Many queueing systems, \textit{e.g.} the security-check areas in airports and ticket offices, can harm the pedestrians in the system and the surrounding environment. 
To address these problems, we constructed a simulation model (Fig. \ref{fig:general floor model}) and surveyed performance measures such as the use ratios, the transit times of the service windows, and the entrance block rate. 

The floor configuration is as follows. 
There is one entrance and $n_s$ service windows. 
The entrance is located at $e$ with the left most  coordinate as one. 
$j$ represents the index of the service windows. 
The intervals between the adjacent service windows are described by $l$. 
The perpendicular distances from the entrance cells to the service windows are set to $L$. 

The arriving agents are injected to the entrance depicted by the red arrow. 
The interval of the arrival follows a log-normal distribution $f_a\left( m_a,\,s_a\right)$, with its mean and standard deviation notated as $m_a$ and $s_a$, respectively. 
Each service window also has a service time distribution $f_s^{(j)}\left( m_s^{(j)},\,s_s^{(j)}\right)$ with a mean, $m_s^{(j)}$, and a standard deviation, $s_s^{(j)}$. 
Both inter-arrival and service times are assumed to follow log-normal distributions. 
An exponential distribution is often assumed for the inter-arrival and service times in basic queueing models. 
Log-normal distributions were introduced instead due to their eligibility in queueing theory, as mentioned in Refs. \cite{Senderovich2015,Yang2016,Brown2005}. 
Discrete inter-arrival and service times were calculated in our simulation as described in Appendix \ref{app:Calculation method for time from a log-normal distribution}. 

The model is based on TASEP with discrete time and parallel updates. 
The update rule of the simulation is configured as follows. 
First, agents stochastically arrive at the entrance of the modeled floor. 
If the entrance cell is empty and a queue in front of the entrance cell is not formed, the agent who arrives can enter the entrance cell. 
Otherwise, the agent has to stand at the end of the queue. 
Next, the agent chooses the target service window by comparing the lengths of the queues and the distances to the service windows. 
Then, they proceed with a probability $p$, reach the target service window, receive service, and finally leave the floor. 

\vspace*{4pt}

\subsection{Window choice}
\label{sec:Model | Window choice}
We integrated two pedestrian tendencies to a method of service window choice for an agent in our model. 
Some pedestrians may prefer to head for less crowded service windows because they want to shorten their waiting times. 
Others may choose closer service windows because they want to shorten their walking times and minimize their possible fatigue. 
In this article, both preferences were incorporated into our model to form an equation that represents the probability of choosing the $j$th service window as follows: 
\begin{equation}
P_j = \frac{1}{Z} \exp\left(-k_N \frac{N_j-E\left[N\right]}{\sqrt{V\left[N\right]}} -k_D \frac{D_j-E\left[D\right]}{\sqrt{V\left[D\right]}}\right), \label{eq:probability of window choice}
\end{equation}
where $Z$ is the normalization constant, $N_j$ is the number of agents heading to the $j$th service window, and $D_j$ is the distance from the entrance cell located at $e$ to the $j$th service window. 
The parameter $k_N$ is the intensity of the influence of the number of agents, $N_j$, on the choice of service windows. 
When there are many pedestrians or obstacles on the floor of a queueing system, the views of newly arrived pedestrians are blocked and they have difficulties in judging which service window is less congested. 
In such a case, staff's guidance can ameliorate pedestrians' choice. 
Guidance in a real situation corresponds to the upward shift of the parameter $k_N$ in our model. 
The parameter $k_D$ is the intensity of the influence of the distance $D_j$. 
$E\left[X\right]$ and $V\left[X\right]$ are the mean and the variance, respectively, of the series $\{X_j\}$ pertaining to $j$. 

\begin{figure}[tbh]
	\centerline{\includegraphics[width=0.8\hsize]{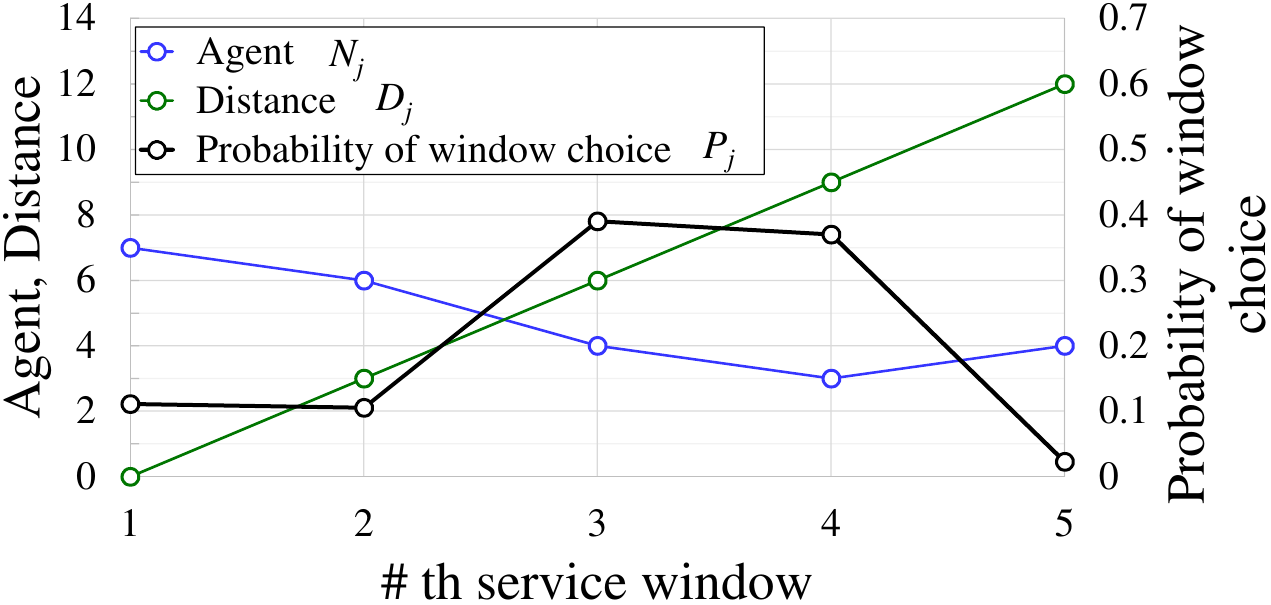}}
	\begin{center}
	\caption{Number of agents $N_j$, the distance $D_j$ and the probability of the service window choice $P_j$ in Fig. \ref{fig:general floor model}. The left vertical axis represents $N_j$ and $D_j$, while the right one represents $P_j$. The service window choice parameters are $k_N=k_D=2$. The other parameters of the floor are the same as that shown in Fig. \ref{fig:general floor model}.}
	\label{fig:window choice example}
	\end{center}
\end{figure}

The number of agents $N_j$, the distance $D_j$ and the probability of the service window choice $P_j$ in Fig. \ref{fig:general floor model} are depicted in Fig. \ref{fig:window choice example}. 
We confirmed that the choice probability for the 3rd service window is larger than that for the 5th one due to its shorter distance for the same number of agents. 
Even though the distance is shortest for the 1st service window, it is not popular because it is chosen by preceding agents. 
 
Because actual pedestrians should proceed to less crowded or nearer service windows, it is reasonable to consider $k_N>0$ and $k_D>0$ for the probability of the service window choice in our model. 
In this paper, the combination of these two parameters is called the `strategy' of the window choice. 

\vspace*{4pt}
\vspace*{4pt}

\subsection{Performance measure}
\label{sec:Model | Performance measure}
In this article, we constructed three performance measures for the model: the use ratio, the transit time, and the entrance block rate. 
The use ratio is the proportion of the number of agents passing through each service window to the total number of agents. 
The transit times are the total times that the agents spend passing through the floor in our simulation, \textit{i.e.}, the total time waiting in the queue ahead of the entrance, walking the aisle to the service window, waiting in the queue in front of the service window, and receiving service. 
The entrance block rate is the rate at which the queues from the service windows extend to the entrance and impede the agents from proceeding.

\subsection{Reference model and strategy}
\label{sec:Model | Reference model and strategy}
\begin{figure}[thb]
	\centerline{\includegraphics[width=9cm]{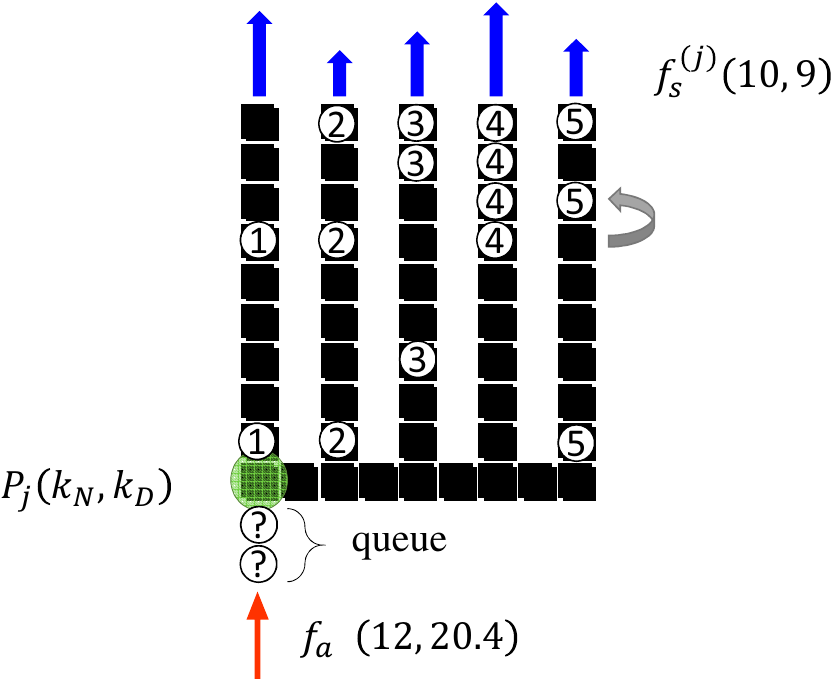}}
	\begin{center}
	\caption{Example state of our simulation using the reference model. The parameters are set as shown in Table \ref{tbl:Reference model parameter}.}
	\label{fig:reference model}
	\end{center}
\end{figure}

All parameters in this model are summarized in Table \ref{tbl:Reference model parameter}. 
We examined the parameters that have a strong effect on the model. 
Owing to the large number of parameters, a reference model was created (Fig. \ref{fig:reference model}) and the influence of each parameter was investigated in comparison to it. 
For the distributions of both the inter-arrival and service times, the coefficients of variation were based on empirical data specified in Ref. \cite{Tanaka2016}.
We set the reference value of the mean of the inter-arrival time to be larger than the service time of the entire floor to achieve a queueing theory stationary state, that is, the waiting times and the number of waiting agents converge. 
For simplicity, the hopping probability $p=1.0$.

Four types of window choice strategies were considered: $R$ indicates a random strategy $(k_N=0,\,k_D=0)$, $N$ indicates a preference for less agents $(k_N=5,\,k_D=0)$, $D$ indicates a preference for shorter distances $(k_N=0,\,k_D=5)$ and $B$ indicates a balance between the two $(k_N=5,\,k_D=5)$. 
We explored the impact of preferences to the numbers of agents and the distances to service windows on the performance measures, such as the transit time and the entrance block rate, by comparing strategies $N$ and $D$. 
The effects of strategies $R$ and $B$ were also assessed. 
Even though $k_N = k_D$ in both strategies $R$ and $B$, agents with strategy $R$ are indifferent to the floor situation while the choices of agents with strategy $B$ depend on both the number of other agents and the distances. 

\begin{table}[t]
\begin{center}
\caption{Parameters of the reference model}
\label{tbl:Reference model parameter}
\vspace*{4pt}
\begin{tabular}{llcc}
\hline
& Parameter				& Symbol 		& Reference value \\
\hline \hline
\multirow{4}{*}{Floor}& Window interval	& $l$ 	& 2 \\
& Floor length			& $L$ 		& 10 \\
& Entrance location	& $e$ 		& 1 \\
& Hopping rate			& $p$ 		& 1 \\
\hline
\multirow{2}{*}{Arrival}& Mean of inter-arrival time & $m_a$ & 12 \\
& Std. of inter-arrival time & $s_a$ & 20 \\ 
\hline
& Number of service windows		& $n_s$			& 5 \\
Service & Mean of service time & $m_s^{(j)}$ 	& 50 \\
& Std. of service time & $s_s^{(j)}$ & 45 \\
\hline
\multirow{2}{*}{Window choice} & Number of agents  & $k_N$ 	& \multirow{2}{*}{$R:\left(\hspace{-.4em}\begin{array}{c} 0 \\ 0 \\ \end{array}\hspace{-.4em}\right)\, N:\left(\hspace{-.4em}\begin{array}{c} 5 \\ 0 \\ \end{array}\hspace{-.4em}\right)\, D:\left(\hspace{-.4em}\begin{array}{c} 0 \\ 5 \\ \end{array}\hspace{-.4em}\right)\, B:\left(\hspace{-.4em}\begin{array}{c} 5 \\ 5 \\ \end{array}\hspace{-.4em}\right)$} \\
& Distance 				& $k_D$ 		& , \hspace{12.3mm} , \hspace{12.3mm} , \hspace{0mm} \\
\hline
\end{tabular} 
\end{center}
\end{table}
\section{Simulation results}
\label{sec:Simulation results}
Thus far, we have described the details of our model, that is, the floor configuration, method of service window choice, performance measures, reference model and choice strategies. 
Hereafter, we provide the results of our simulations. 
First, we investigate the effect of the window choice strategy on the use ratio and the transit time in Sec. \ref{sec:Simulation results | Window choice strategy}. 
Then, the average transit time and entrance block rate are analyzed by altering each parameter of the model. 
We investigate the state where 500 agents pass through the floor after the first 10,000 steps. 
This trial was repeated 10,000 times for each simulation condition. 
The error bars in the figures represent the standard deviation of the results of the 10,000 trials. 

\subsection{Window choice strategy}
\label{sec:Simulation results | Window choice strategy}
\begin{figure}[htb]
	\begin{minipage}{0.5\hsize}
		\begin{center}
			\includegraphics[width=0.95\hsize]{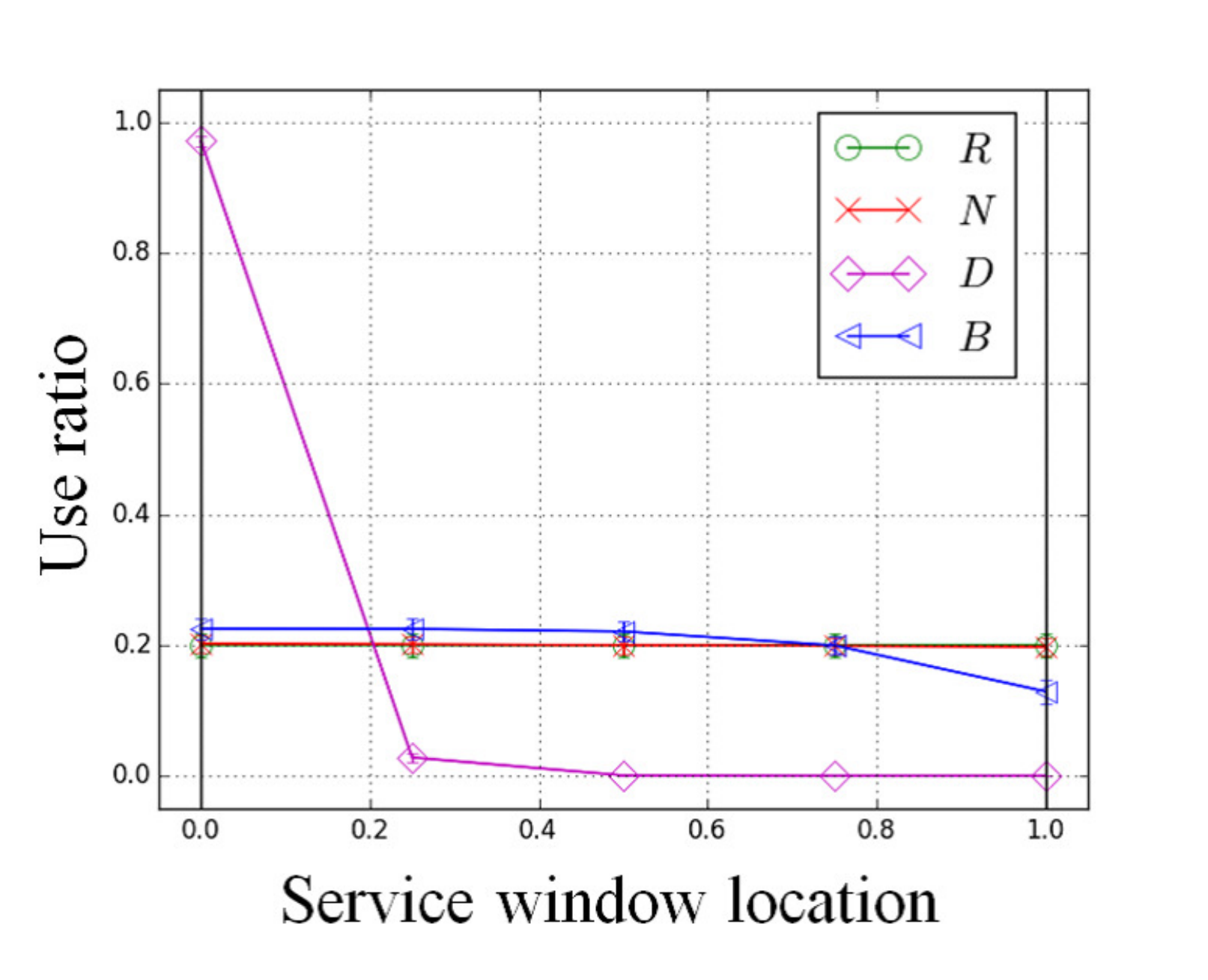}
			\hspace{2cm} (a)
		\end{center}
	\end{minipage}
	\begin{minipage}{0.5\hsize}
		\begin{center}
			\includegraphics[width=0.95\hsize]{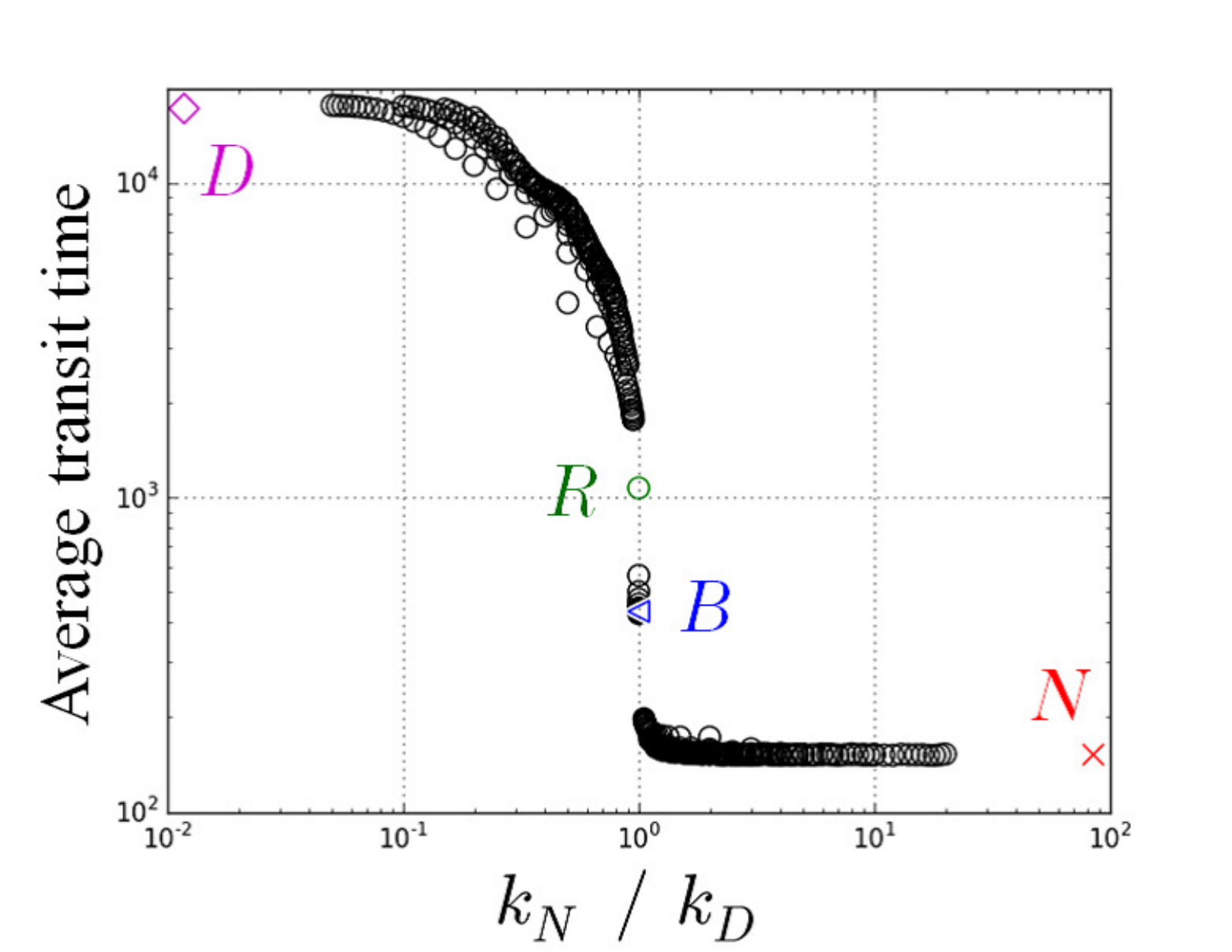}
			\hspace{2cm} (b)
		\end{center}
	\end{minipage}
	\caption{Characteristics of the four strategies for the reference model: (a) use ratio of the service windows and (b) average transit time as a function of $k_N/k_D$. The transit times of strategies $R$, $N$, and $D$ are plotted at $k_N / k_D = 1$, $10^{1.93}$ and $10^{-1.93}$, respectively.}
	\label{fig:use_ratio_time_reference}
\end{figure}

The use ratio and the transit time of the service windows were analyzed for the four strategies described in Table \ref{tbl:Reference model parameter}. 
In Fig. \ref{fig:use_ratio_time_reference} (a), which shows the use ratio, we see results consistent with our intuition for the three strategies $R$, $N$ and $D$, \textit{i.e.}, the service windows were selected equally for strategies $R$ and $N$ while the ones closer to the entrance were well used for strategy $D$. 
Strategy $B$ offered a blend between these strategies and the agents declined gradually with distance from the entrance. 
This can be attributed to the tendency for agents to head for closer service windows unless the number of waiting agents there becomes too large.  

The transit times for strategy $D$ became large (Fig. \ref{fig:use_ratio_time_reference} (b)). 
This resulted from congestion near the entrance due to the preference for shorter distances. 
Because extraordinary large values were calculated for strategy $D$ and its practical applications are limited, we concentrated on investigating strategies $R$, $N$, and $B$ in the following sections. 

The average transit times of all service windows were calculated for the strategy parameters $(k_N,\,k_D)$ varying over the range of $\left[0.5,\,10\right]\times\left[0.5,\,10\right]$ as a function of their ratios $k_N/k_D$ (Fig. \ref{fig:use_ratio_time_reference} (b)). 
The times for the above mentioned strategies are also shown. 
Note that the results of strategies $R$, $N$, and $D$ are plotted at $k_N / k_D = 1$, $10^{1.93}$ and $10^{-1.93}$, respectively, for convenience. 
When $k_N \ll k_D$, the average transit time becomes large. 
Meanwhile, when $k_N \gg k_D$, the average transit time converges to a small value. 
For the blended situation the average transit time grew steeply with decreasing values of $k_N/k_D$. 
Therefore, we can safely focus on the three strategies $R$, $N$, and $D$ that fell into each range. 
This results from the linearity of the strategy parameters in the window choice definition in Eq. (\ref{eq:probability of window choice}). 

The average transit times are arranged from strategies $R$ to $B$ to $N$ in descending order.
Regarding strategies $N$ and $R$, agents for both could be distributed equally from a long-term perspective. 
However, their transit times differ from each other. 
In strategy $N$, agents covered the queues with fewer agents immediately when there was a bias in the number of agents toward the service windows. 
By contrast, in strategy $R$, agents selected their windows at random. 
Therefore, congestion lasted longer with strategy $R$. 
Even though this equality mechanism for the number of agents was also valid for strategy $B$, the transit times increased compared to those with strategy $N$ because more agents were willing to pass the closer service windows.  

\vspace*{4pt}

\subsection{Utility rate}
\label{sec:Simulation results | Utility rate}
\begin{figure}[th]
	\begin{minipage}{0.5\hsize}
		\begin{center}
			\includegraphics[width=0.95\hsize]{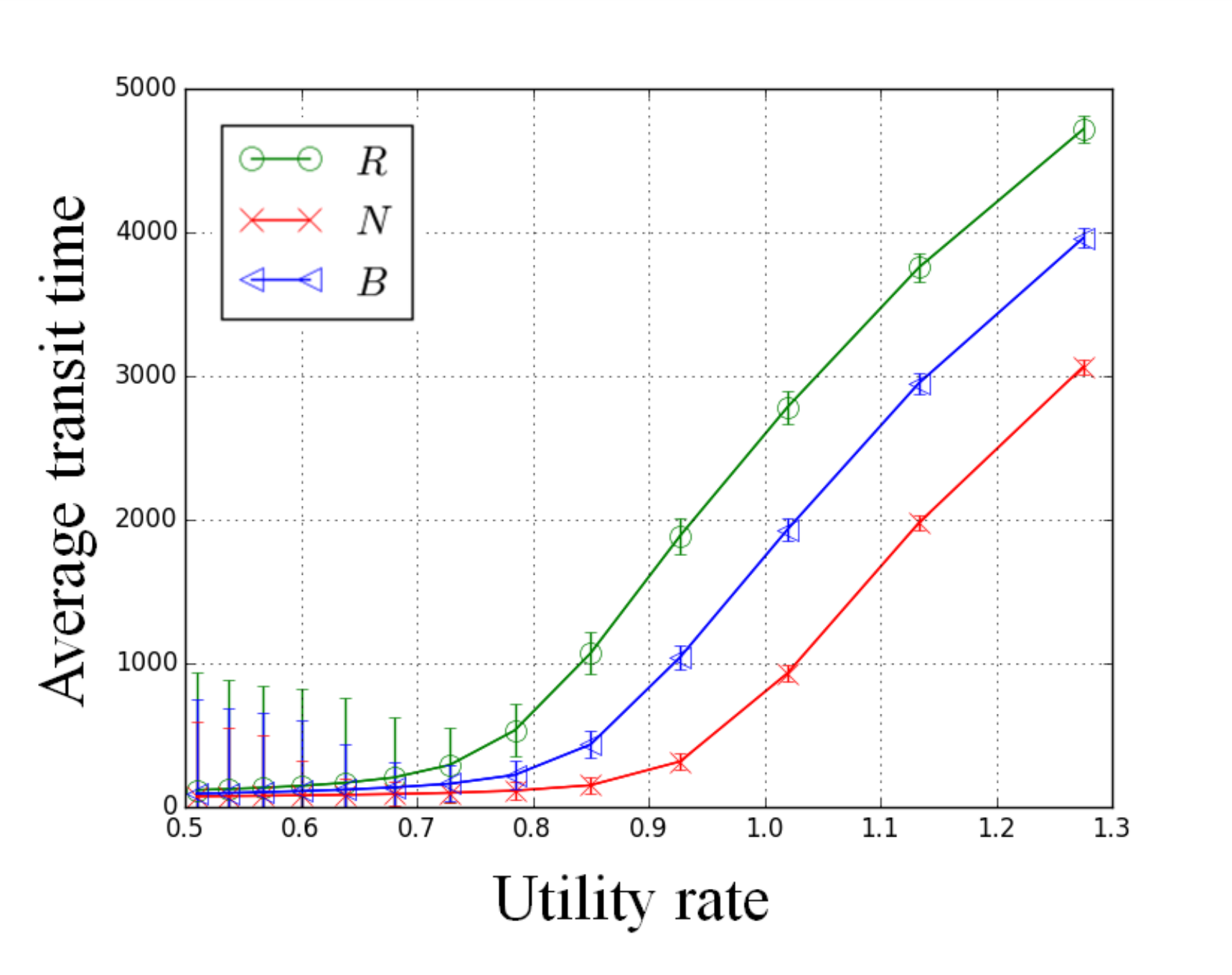}
			\hspace{2cm} (a)
		\end{center}
	\end{minipage}
	\begin{minipage}{0.5\hsize}
		\begin{center}
			\includegraphics[width=0.95\hsize]{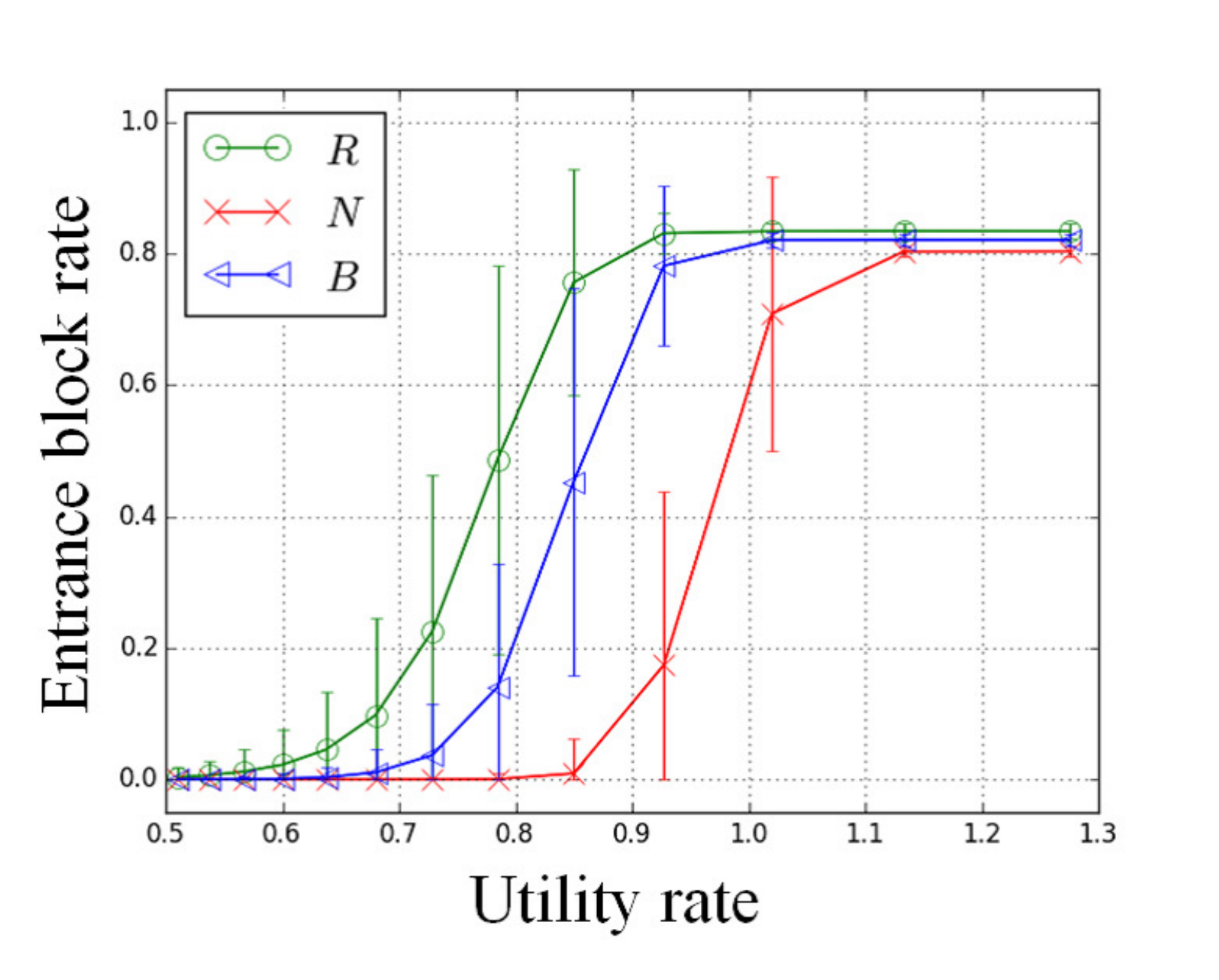}
			\hspace{2cm} (b)
		\end{center}
	\end{minipage}
	\caption{Average transit time (a) and entrance block rate (b) as a function of utility rate given by $51/5m_a$ using the mean of inter-arrival time $m_a$. $m_a$ were set from 8 to 20 at the interval of 1.}
	\label{fig:average_time_entrance_block_mwte}
\end{figure}

The average transit times and entrance block rates as a function of the utility rate are depicted in Fig. \ref{fig:average_time_entrance_block_mwte}. 
The utility rate is defined as $(m_s^{(j)}+1)/m_a$ \protect\linebreak

\noindent $=51/5m_a$ as in the exclusive queueing model \cite{Yanagisawa2010}. 
The average transit times increase with the utility rate, while the thresholds of abrupt surges are arranged from strategies $R$ to $B$ to $N$. 
Figure \ref{fig:average_time_entrance_block_mwte} (b) suggests why this happened, \textit{i.e.}, the rising entrance block rates caused the increase in the average transit times. 
The increases in the average transit times became steep for entrance block rates greater than 0.2. 

A utility rate greater than 1.0 indicates the arrival of more agents than the service capacity, and this condition should lead to a diversion in the number of agents and their transit times. 
However, the average transit time was relatively small for a utility rate of slightly more than 1.0 for strategy $N$ as shown in Fig. \ref{fig:average_time_entrance_block_mwte} (a). 
This is because with strategy $N$, agents' window choices were strongly influenced by the number of agents toward the service windows and the agents were nearly evenly distributed. 
When the simulation finished after passing the target of 500 agents, the remaining agents could be accommodated within the floor cells and the entrance cell was not likely to hinder subsequent agents by being occupied. 
Therefore, the transit time for strategy $N$ with a utility rate of slightly more than 1.0 did not have a large value during the simulation times. 
Even though the diversion was not clearly observed for strategy $N$ under the simulation condition, it can reach an obvious diversion with additional simulation steps, as seen in the other strategies. 
For strategies $R$ and $B$, more frequent choices of longer queues resulted in quick diversions.

\subsection{Window interval: $l$}
\label{sec:Simulation results | Window interval}
\begin{figure}[htb]
	\begin{minipage}{0.5\hsize}
		\begin{center}
			\includegraphics[width=0.95\hsize]{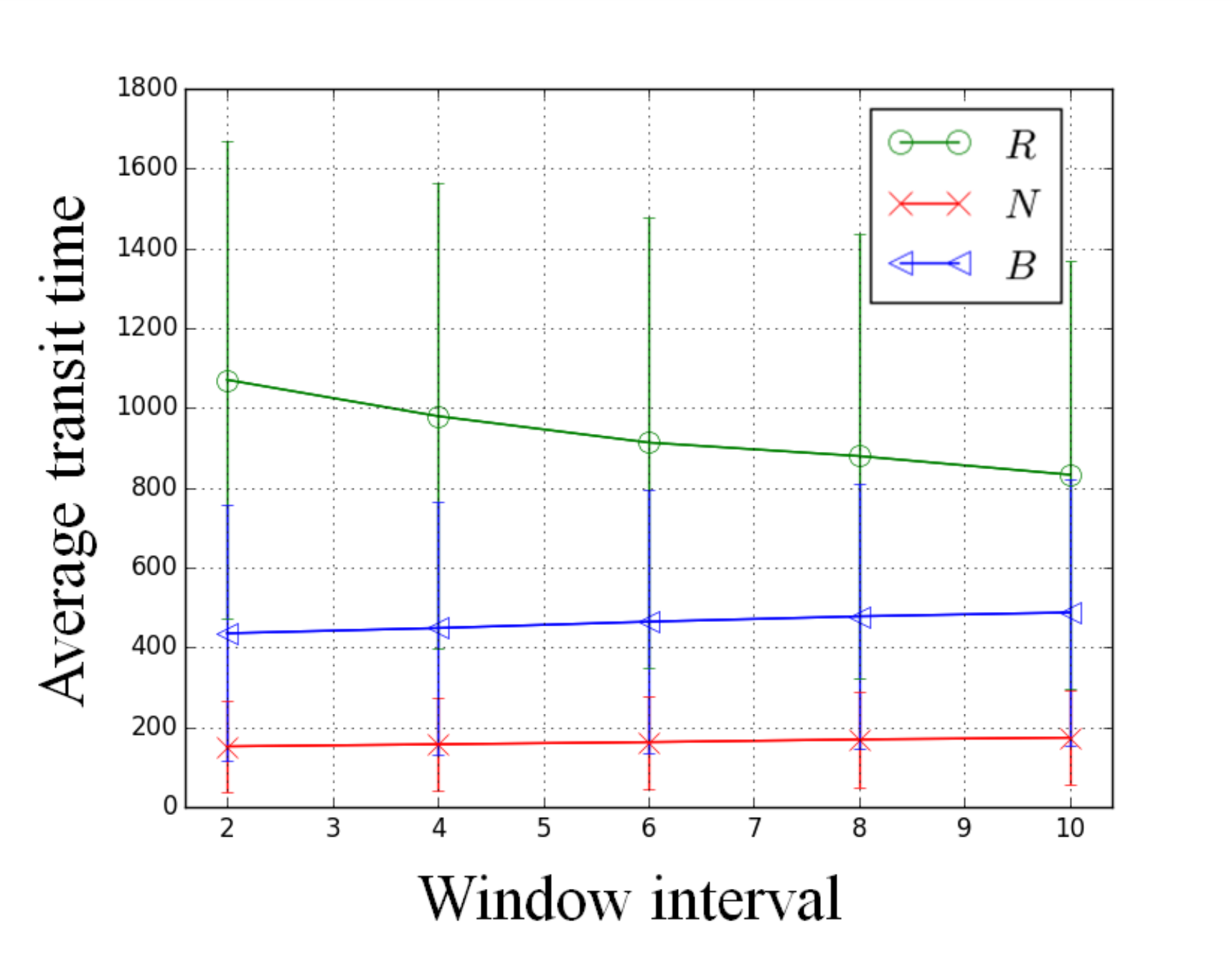}
			\hspace{2cm} (a)
		\end{center}
	\end{minipage}
	\begin{minipage}{0.5\hsize}
		\begin{center}
			\includegraphics[width=0.95\hsize]{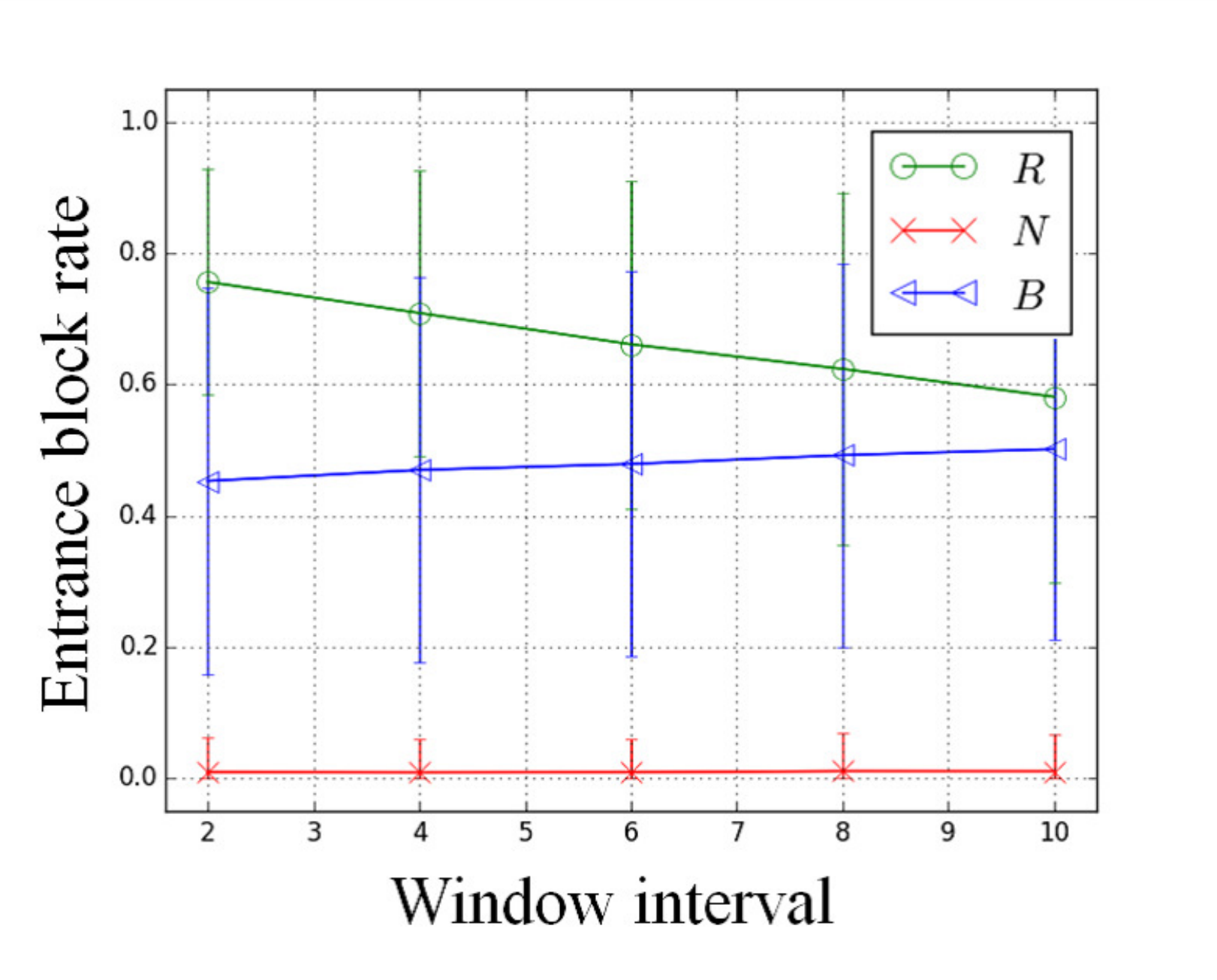}
			\hspace{2cm} (b)
		\end{center}
	\end{minipage}
	\caption{(a) Average transit time and (b) entrance block rate as a function of the window interval $l$, from 2 to 10 with an interval of 2.}
	\label{fig:average_time_entrance_block_l}
\end{figure}

The influence of the intervals between service windows, $l$, was explored (Fig. \ref{fig:average_time_entrance_block_l}). 
Both the average transit time and the entrance block rate share the same tendency, that is, for strategies $N$ and $B$, their values slightly increased, while for strategy $R$, their values decreased. 
The increase in the transit time was caused by the greater walking distance for the agents to the service windows. 
Even though the distance affected the transit time and block rate in a similar fashion for strategy $R$, the reduction of the congestion had a greater influence. 
The probability of the entrance being encumbered by longer queues from remote service windows became small with larger $l$ (Fig. \ref{fig:average_time_entrance_block_l} (b)), causing the average transit time to decrease (Fig. \ref{fig:average_time_entrance_block_l} (a)).

\vspace*{4pt}

\subsection{Floor length: $L$}
\label{sec:Simulation results | Floor length}
\begin{figure}[htb]
	\begin{minipage}{0.5\hsize}
		\begin{center}
			\includegraphics[width=0.95\hsize]{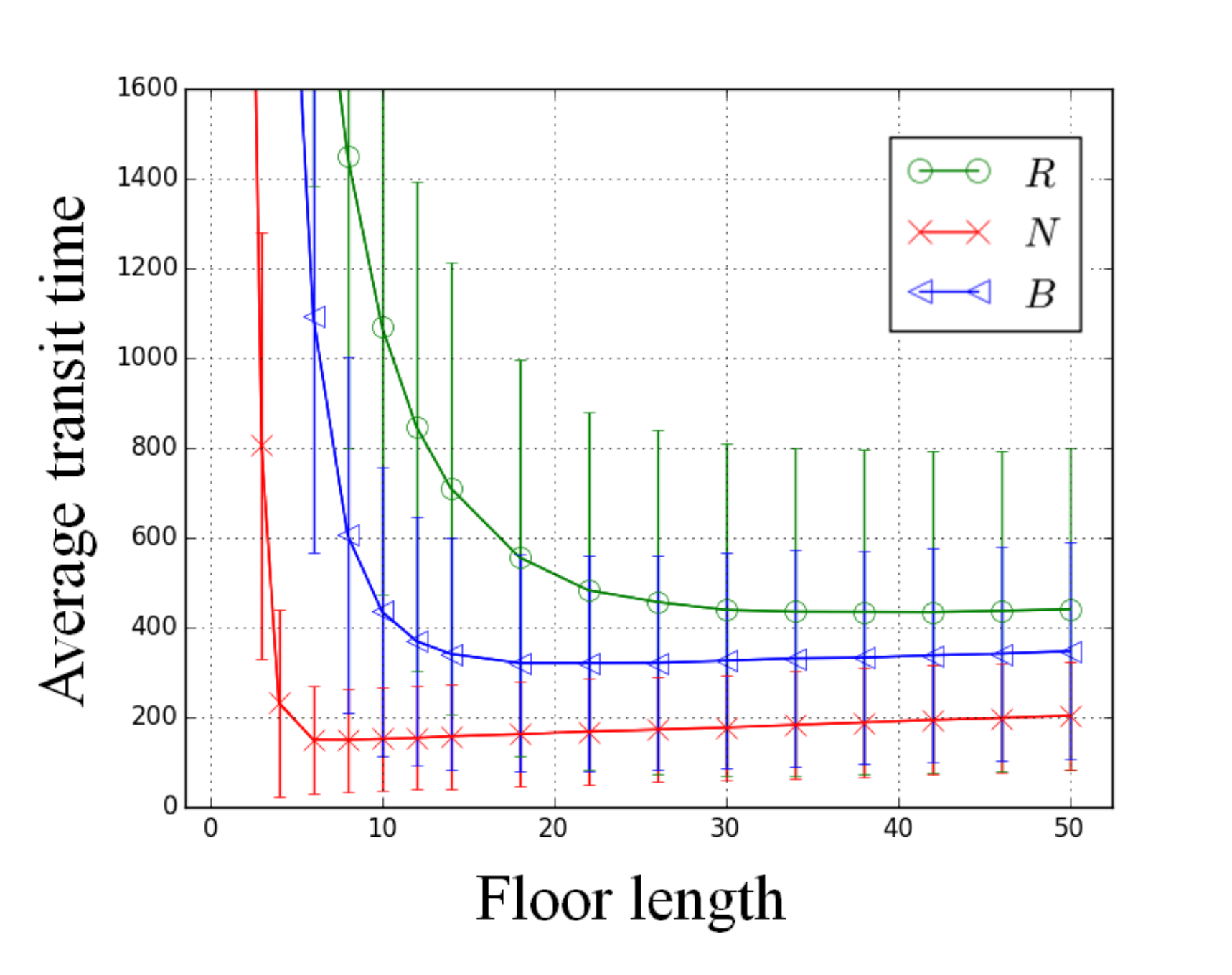}
			\hspace{2cm} (a)
		\end{center}
	\end{minipage}
	\begin{minipage}{0.5\hsize}
		\begin{center}
			\includegraphics[width=0.95\hsize]{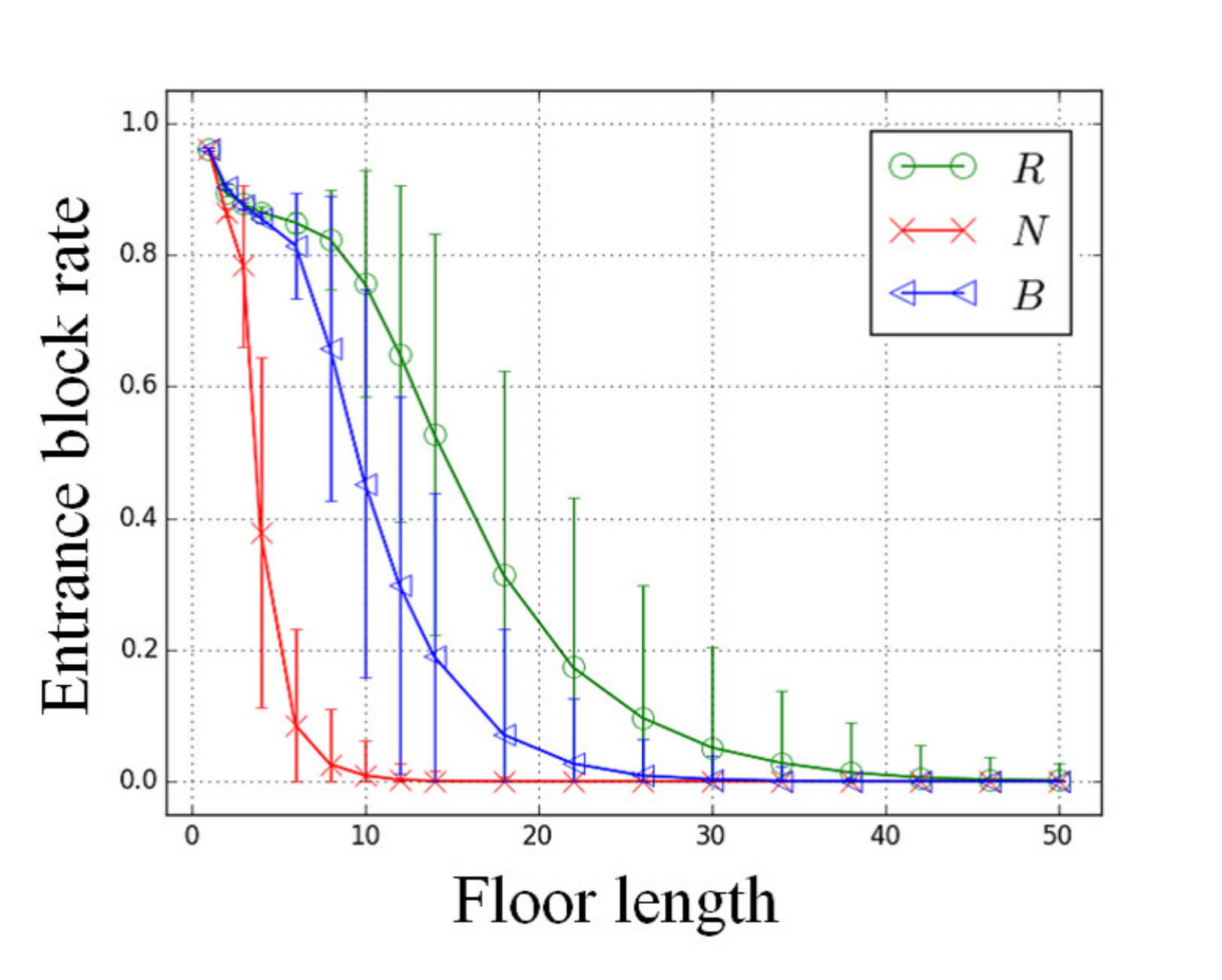}
			\hspace{2cm} (b)
		\end{center}
	\end{minipage}
	\caption{(a) Average transit time and (b) entrance block rate as a function of the floor length, $L$, for $1,\,2,\,3,\,4,\,6,\,\cdots ,\,14,\,18,\,\cdots ,\,50$.}
	\label{fig:average_time_entrance_block_fl}
\end{figure}

In this section, the effects of the floor length $L$ are investigated. 
The parameter $L$ can be interpreted as the maximum waiting agent capacity for the service windows. 
The transit times decrease as the floor length $L$ increases when $L$ is small (Fig. \ref{fig:average_time_entrance_block_fl} (a)). 
Having a small $L$ deprives the agents of space to wait, so that queues from the service windows easily blocked the entrance cell (Fig. \ref{fig:average_time_entrance_block_fl} (b)). 
Therefore, having more space for agents to wait in prevented the queues from encumbering the entrance. 
However, the average transit time starts to increase when $L$ is further increased. 
Having a large distance to walk accounts for this phenomenon; this is similar to the effect seen for the window intervals $l$. 
Consequently, a minimum transit time exists for a floor length $L$, as depicted in Fig. \ref{fig:average_time_entrance_block_fl} (a). 
The minimum values for strategies $R$, $N$, and $B$ were achieved when the floor lengths were 6, 18, and 42, respectively.

\vspace*{4pt}

\subsection{Number of service windows: $n_s$}
\label{sec:Simulation results | Number of service windows}
To examine the average transit times and entrance block rates for different numbers of service windows, we assume that the mean service time of each service window increases proportionally to the number of service windows owing to the coherence of the total service rate: 
\begin{equation}
m_s^{(j)}
=50\cdot\frac{n_s}{5}=10n_s.
\end{equation}

\begin{figure}[thb]
	\begin{minipage}{0.5\hsize}
		\begin{center}
			\includegraphics[width=0.95\hsize]{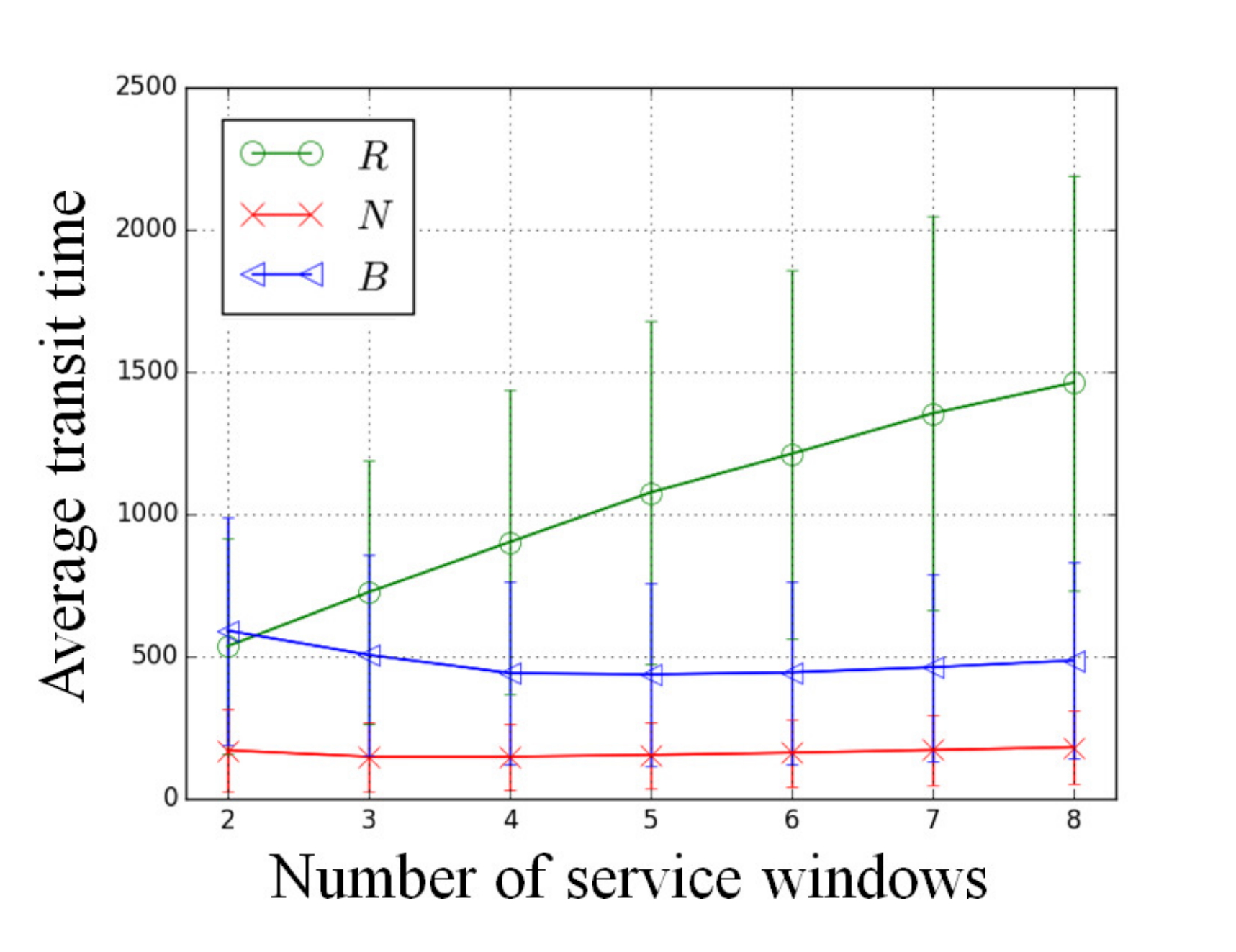}
			\hspace{2cm} (a)
		\end{center}
	\end{minipage}
	\begin{minipage}{0.5\hsize}
		\begin{center}
			\includegraphics[width=0.95\hsize]{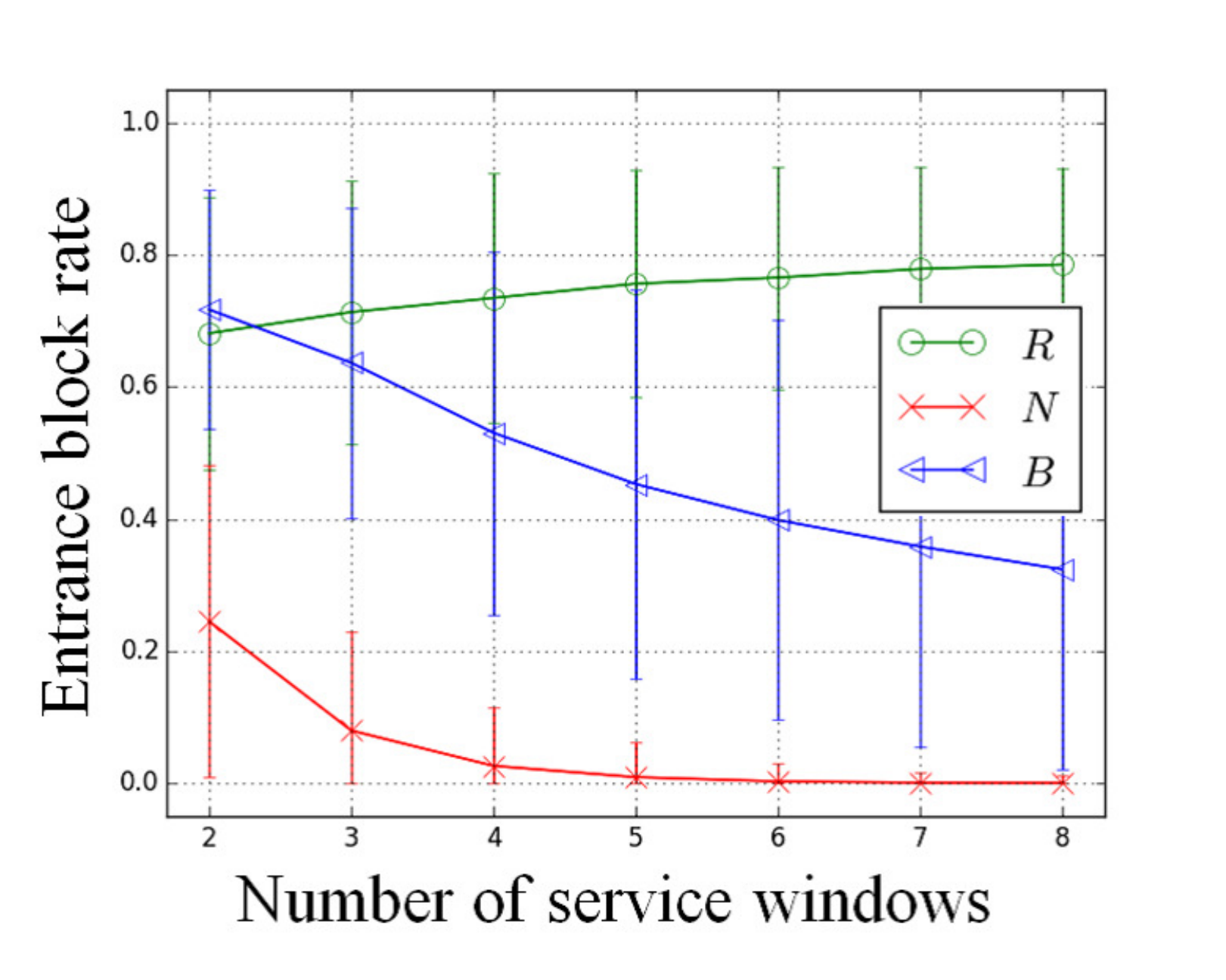}
			\hspace{2cm} (b)
		\end{center}
	\end{minipage}
	\caption{(a) Average transit time and (b) entrance block rate as a function of the number of service windows, $n_s$, from 2 to 8 with an interval of 1.}
	\label{fig:average_time_entrance_block_s}
\end{figure}

The results of varying the number of service windows are depicted in Fig. \ref{fig:average_time_entrance_block_s}. 
The average transit time for strategy $R$ increased, while those for the other strategies are downwardly convex (Fig. \ref{fig:average_time_entrance_block_s} (a)).  
The primary factor affecting the curves for strategies $N$ and $B$ is the extended walking distance. 
As the number of service windows $n_s$ increases, the walking time increases with the walking distance as explained in Secs. \ref{sec:Simulation results | Window interval} and \ref{sec:Simulation results | Floor length}. 
Conversely, greater distance prevents queues from reaching the entrance and decreases the entrance block rate (Fig. \ref{fig:average_time_entrance_block_s} (b)). 
These effects of distance have a trade-off relationship for the average transit time. 
The minimum values for strategies $N$ and $B$ were reached when the numbers of service windows were 3 and 5. 

\vspace*{4pt}
\vspace*{4pt}

The increase for strategy $R$ is expounded as follows. 
We considered the utility rate for the $j$th service window as follows: 
\begin{equation}
\rho^{(j)} := \frac{P_j}{m_a} \bigg/ \frac{1}{m_s^{(j)}+1}.
\end{equation}
According to the convergence condition for the $j$th service window, $\rho^{(j)}<1$, the critical choice probability 
\begin{equation}
P_{\mathrm{cr}} := \frac{m_a}{m_s^{(j)}+1} = \frac{m_a}{10n_s + 1}
\end{equation}
can be calculated. 
Because the critical choice probability $P_{\mathrm{cr}}$ decreases with increases in the number of service windows, less agents can choose to go to the same $j$th service window without a queue diversion. 
Here we define $X_j$ as the number of events where the $j$th service window is chosen by agents. 
The vector $\mathbf{X}=(X_1,\,X_2,\,\cdots ,\,X_{n_s})$ follows a multinomial distribution with a size of 500 and a probability of $1/n_s$ because, in strategy $R$, the 500 agents choose the service windows with the same probability. 
If $X_j$ is smaller than the critical number of agents who chose the $j$th service window, that is, 
\begin{equation}
X_j<500P_{\mathrm{cr}},
\label{eq:Xj<500Pcr}
\end{equation}
we consider a divergence as having not occurred for the $j$th service window. 
Because our simulations run with only 500 agents, an actual divergence did not occur; however, the transit times became significantly large for agents passing through the service windows that did not satisfy the inequality (Eq. (\ref{eq:Xj<500Pcr})). 
Applying the central limit theorem, 
the convergence condition for the floor: 
\begin{eqnarray}
&&\hspace*{-18pt} Prob \left(\forall j\;\; X_j < 500P_{\mathrm{cr}}\right) \nonumber \\
&&\hspace*{-13pt}\; = \int_{A}\frac{1}{\left(\sqrt{2\pi}\right)^{n_s}\sqrt{|\Sigma|}}\exp\left( -\frac{1}{2} \left(\mathbf{X}-500\mathbf{P}_{\mathrm{cr}}\right)^\mathrm{T} \mathbf{\Sigma}^{-1} \left(\mathbf{X}-500\mathbf{P}_{\mathrm{cr}} \right)\right) \mathrm{d}\mathbf{X}
\label{eq:convergence probabilities by multivariate normal distribution}
\end{eqnarray}
was calculated numerically, where $A$ indicates the realm where the convergence occurs, $\mathbf{\Sigma}$ is the variance--covariance matrix of $X_j$ and $\mathbf{X}$ and $\mathbf{P}_{\mathrm{cr}}$ are vectors whose $j$th elements are $X_j$ and $P_{\mathrm{cr}}$, respectively. 

\begin{figure}[h]
	\begin{center}
		\includegraphics[width=0.6\hsize]{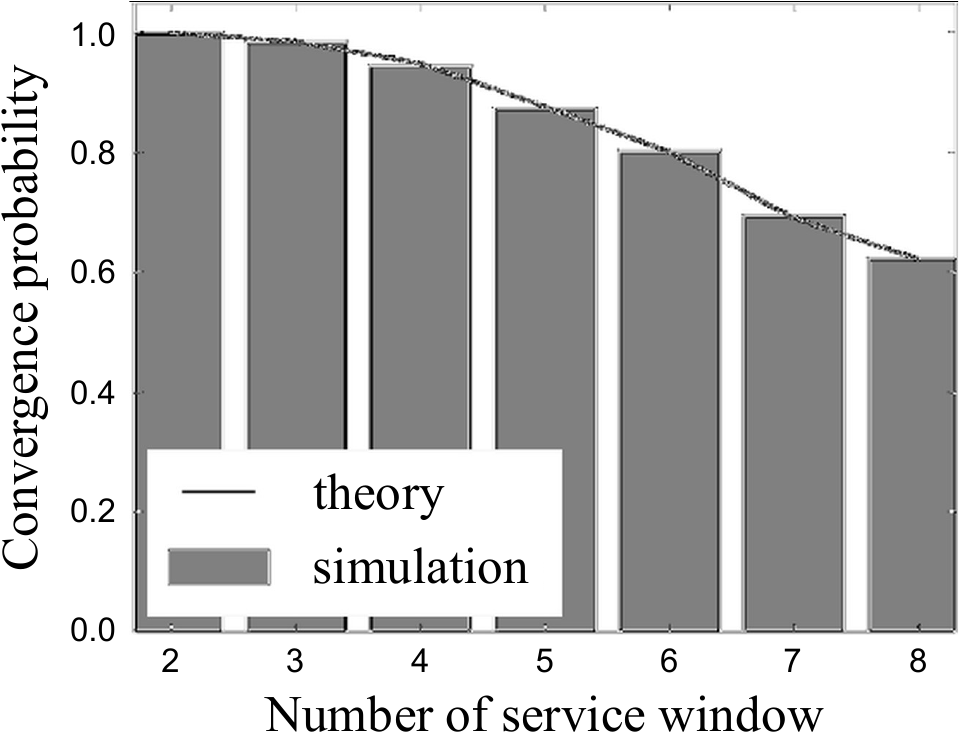}
	\end{center}
	\caption{Convergence probabilities calculated by analytic equations (Eq. (\ref{eq:convergence probabilities by multivariate normal distribution})) (solid line) and those derived by the simulations (gray bars) as a function of the number of service windows $n_s$.}
	\label{fig:convergence probabilities for s}
\end{figure}

This analytic probability is plotted in Fig. \ref{fig:convergence probabilities for s} with the simulation result calculated from the use ratio. 
They agree very well with each other. 
With increasing numbers of service windows $n_s$, convergence becomes less likely to occur. 
Therefore, both the transit time and the entrance block rate for strategy $R$ increased as shown in Fig. \ref{fig:average_time_entrance_block_s}. 

\vspace*{4pt}

\subsection{Entrance location: $e$}
\label{sec:Simulation results | Entrance location}
\begin{figure}[htb]
	\begin{minipage}{0.5\hsize}
		\begin{center}
			\includegraphics[width=0.95\hsize]{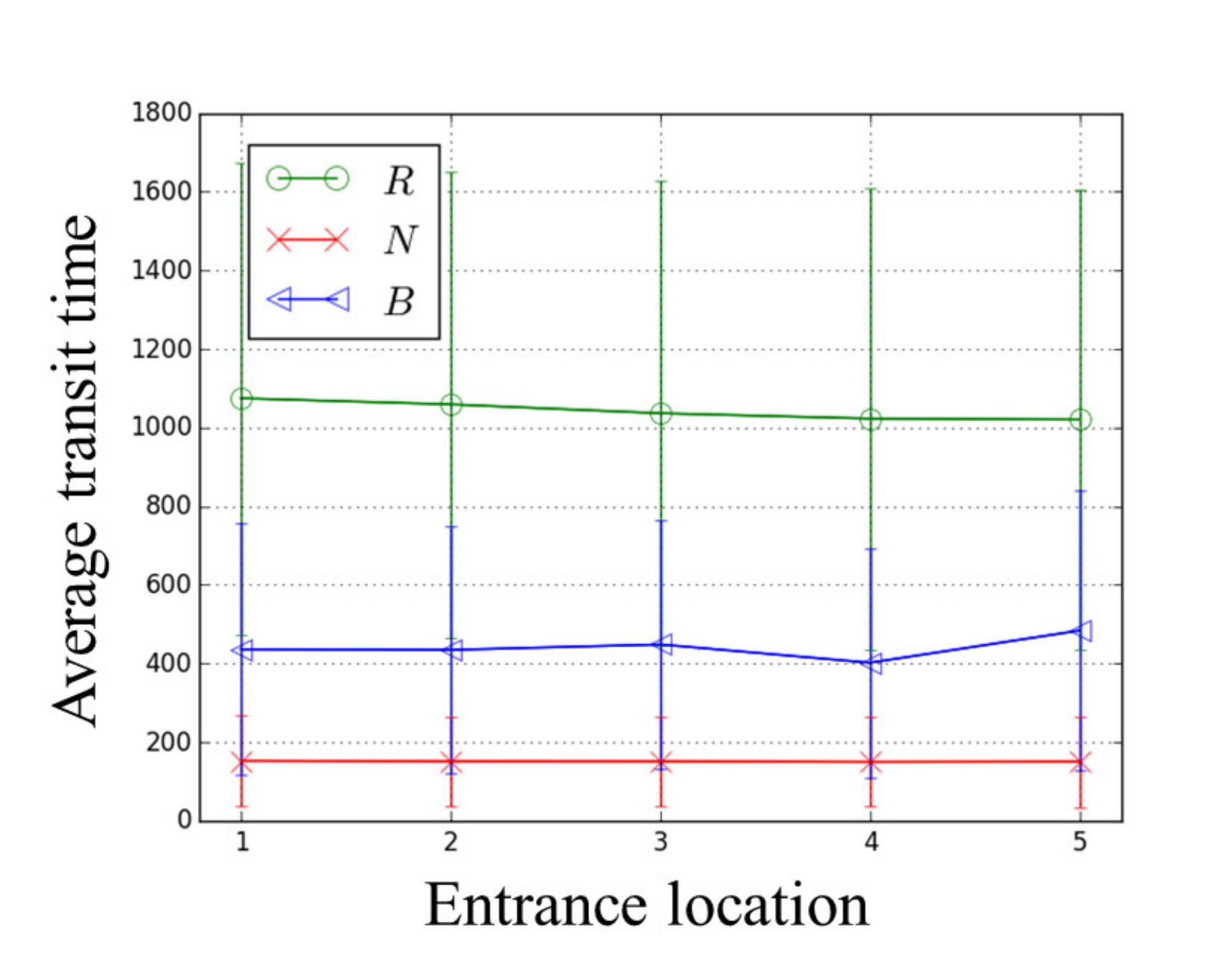}
			\hspace{2cm} (a)
		\end{center}
	\end{minipage}
	\begin{minipage}{0.5\hsize}
		\begin{center}
			\includegraphics[width=0.95\hsize]{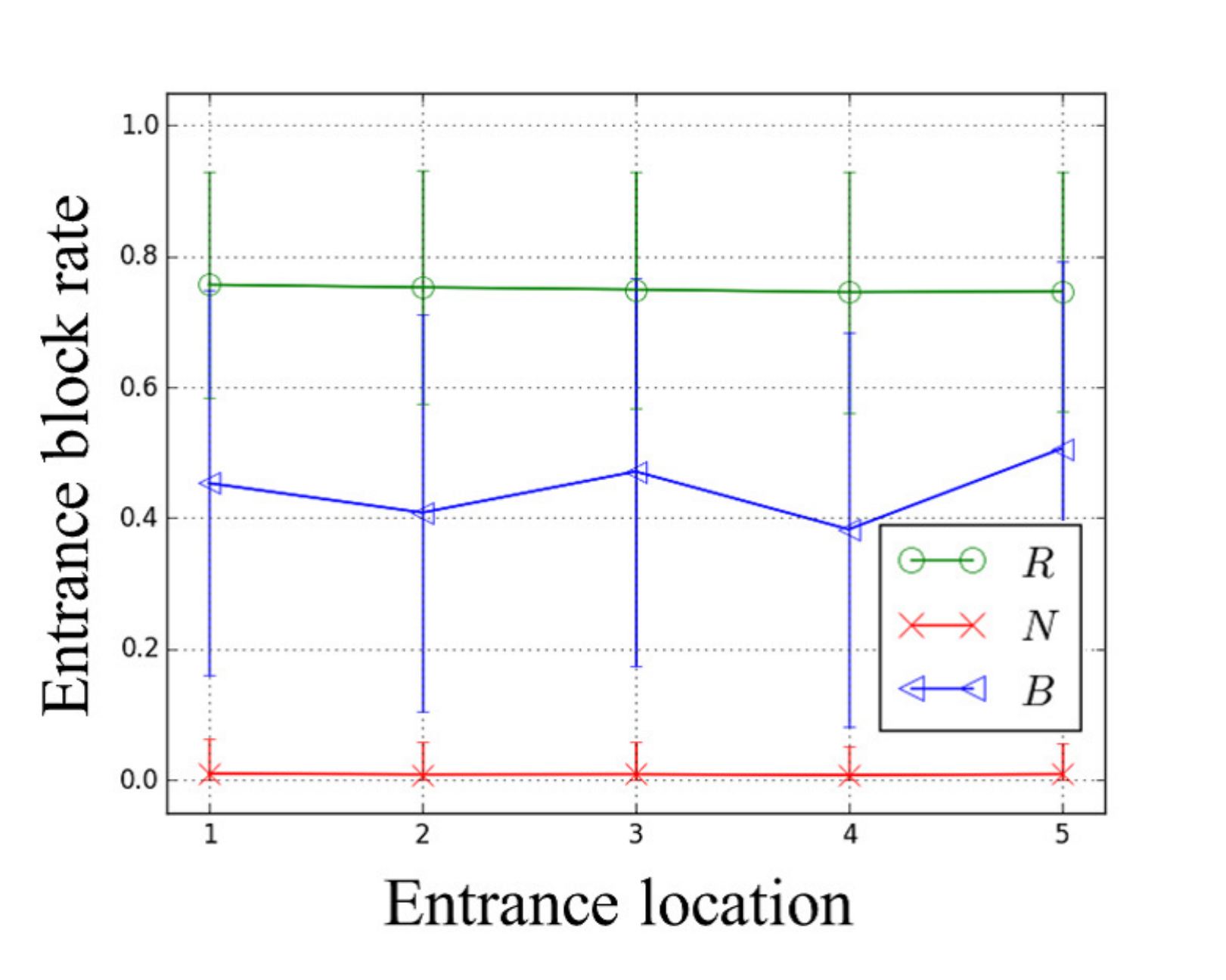}
			\hspace{2cm} (b)
		\end{center}
	\end{minipage}
	\caption{(a) Average transit time and (b) entrance block rate as a function of the entrance location, $e$, from 1 (left edge) to 5 (center). Owing to the symmetry of the reference floor, it is sufficient for the left-half entrance location. }
	\label{fig:average_time_entrance_block_entrance}
\end{figure}

In this section, the influence of the entrance location, $e$, was investigated (Fig. \ref{fig:average_time_entrance_block_entrance}). 
The left-half entrance locations of the floor are sufficient to be plotted because the floor conditions without the entrance location have lateral symmetry. 
Both the average transit time and the entrance block rate exhibit the same trend. 
As the entrance is altered from the edge of the horizontal aisle to the center, for strategy $R$, they slightly decreased; moreover, for strategy $N$, they remained unchanged, and for strategy $B$, they oscillated. 
The slight decreases observed for strategy $R$ were caused mainly by the walking distance. 
The average of the horizontal walking distance from the entrance to all the service windows decreases with the shifting of the entrance location. 

We calculated the oscillations of both the average transit time and the entrance block rate for strategy $B$.
They were attributed to whether the entrance was located in front of a service window ($e=$ odd numbers) or not ($e=$ even numbers). 
In the former situation, the choice probability for the front service window was stronger than that for the two closest service windows in the latter situation. 
Therefore, the entrance block rates for $e=$ odd numbers were higher than that for $e=$ even numbers. 
This mechanism led to the oscillation of the average transit time. 

\vspace*{4pt}

\subsection{Average transit time versus entrance block rate}
\label{sec:Simulation results | Average transit time versus Entrance block rate}
\begin{figure}[h]
	\begin{center}
		\includegraphics[width=0.7\hsize]{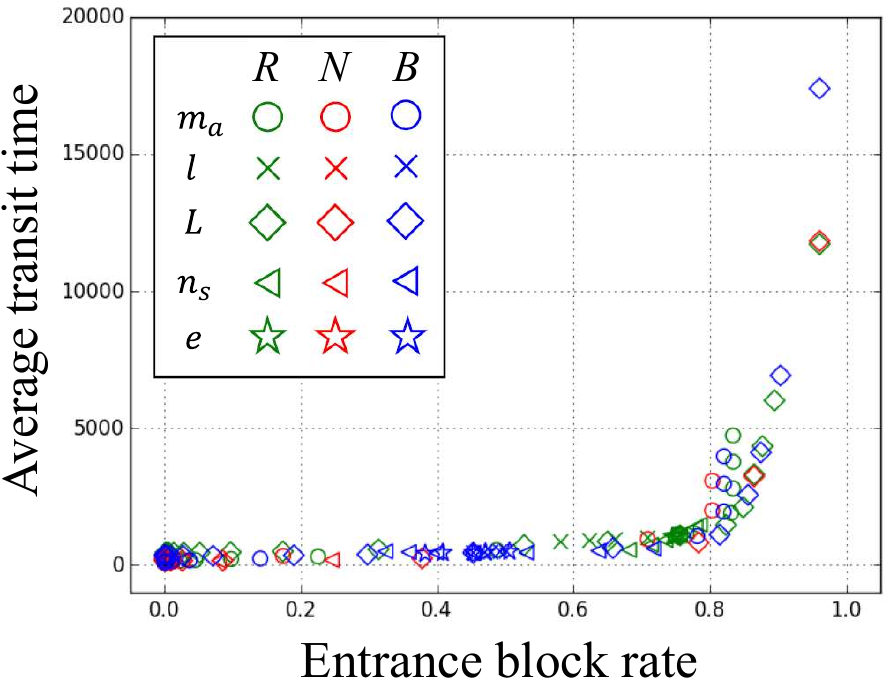}
	\end{center}
	\caption{Relationship between the entrance block rate and the average transit time. Each marker is defined in the figure key. The results in Secs. \ref{sec:Simulation results | Utility rate} ($m_a$), \ref{sec:Simulation results | Window interval} ($l$), \ref{sec:Simulation results | Floor length} ($L$), \ref{sec:Simulation results | Number of service windows} ($n_s$), and \ref{sec:Simulation results | Entrance location} ($e$) are indicated by circles, crosses, diamonds, triangles, and stars, respectively. The colors indicate strategies: $R$ (green), $N$ (red), and $B$ (blue).}
	\label{fig:Average_time_v.s._Entrance_block}
\end{figure}

Finally, we demonstrate the simultaneous representation of the relationship between the average transit times and the entrance block rates explained in Secs. \ref{sec:Simulation results | Utility rate} -- \ref{sec:Simulation results | Entrance location}. 
Figure \ref{fig:Average_time_v.s._Entrance_block} suggests that both characteristics are mutually dependent. 
The average transit time slightly increased for an entrance block rate under 0.8, while it rapidly increased for values over 0.8. 

The service windows have been assumed to be the bottleneck in standard queueing models. 
However, our simulations confirm that the entrance has an indisputable influence on our model. 
Therefore, it is important to investigate the entrance condition in addition to the service windows. 

\vspace*{4pt}

\section{Conclusions}
\label{sec:Conclusions}
In this paper, we analyzed a queueing floor represented by discrete cells including service window choice from the viewpoint of transit times and entrance block rates. 
Agents can perceive the choices of other agents on the floor and the distances to each service window. 
Choices were determined by the agents on a basis of their strategies, which represents the weight they assign to the abovementioned elements, and logit-based probabilities were introduced to demonstrate the choices in our simulation. 

We obtained remarkable results. 
The strategies of the agents exerted a significant influence on the transit times: if the distances to the service windows are considered to be more important by the agents, the transit times become very large, and if the numbers of agents are emphasized more, the transit times become small. 
There are optimal floor lengths for each strategy and an optimal number of service windows for some strategies that minimize the average transit times of agents. 
These results were derived from the trade-off in the distances to the service windows, that is, the trade-off between preventing queues from the service windows from reaching the entrance and the greater walking distance. 
When the entrance is located at center, the walking time decreases; thus, the transit time also becomes small in some window choice strategies. 
However, if the entrance is located in front of a service window, a long queue encumbers the agents entering from the entrance. 
Thus, the transit time increases for other strategies. 
These results can be employed in facility design or floor management. 
In addition, we discussed the relationship between the entrance block rate and the average transit time to find their mutual dependence. 
We propose that the entrance condition should be investigated further in queueing systems in addition to the service window conditions. 
We also examined the influence of the means of the inter-arrival times and the window intervals on the transit time and entrance block rate. 

The core of the proposed model is the selection of the queue as one element of the service window choice. 
We broadened the horizon of the application of exit choice, that is, from exit choice only in the evacuation literature to service window choice in the floor-management literature. 
We adopted logit-based probability as a choice mechanism. 
Our simulation could reproduce results similar to those reported in previous studies. 
Lo et al. \cite{Lo2006} and Fu et al. \cite{Fu2014} revealed that if agents underline the congestion when choosing exits, their evacuation times decrease. 
We derived small average transit times if the agents tended to choose service windows with short queues. 
The shorter queue in our model corresponds to the less congested exit in the models of Lo et al. and Fu et al. 
Therefore, our model using logit-based probability is feasible for addressing congestion issues in queueing systems. 

The proposed model can be applied to a broad range of studies and the management of queueing systems, such as security-check areas in airports and ticket gates in amusement parks. 
As mentioned in Sec. \ref{sec:Introduction}, our choice model concentrated on pedestrian densities around all service windows and the distances to the service windows. 
However, there may remain other factors in criteria of choice of service windows. 
In evacuation studies, the behavior of other pedestrians and the flow of a particular exit, such as an emergency exit, have been found to be influential on exit choice behavior \cite{Augustijn-Beckers2010}. 
Whether such elements are significant for queueing models remains to be studied. 
In addition to the conducted simulation analysis of our model, we have already collected empirical data and evaluated the validity of our model for a special case \cite{Tanaka2016}. 
Our model assumes a single homogeneous strategy of service window choice, however, pedestrians in the real world have their own preferences. 
Therefore, heterogeneous strategies for choosing service windows could be conceived as an extension of our model. 
%

\appendix
\renewcommand{\theequation}{\Alph{section}.\arabic{equation}}
\setcounter{equation}{0}
\renewcommand{\thetable}{\Alph{section}.\arabic{table}}
\setcounter{table}{0}
\renewcommand{\thefigure}{\Alph{section}.\arabic{figure}}
\setcounter{figure}{0}
\section{Calculation method for time from a log-normal distribution}
\label{app:Calculation method for time from a log-normal distribution}
The calculations were conducted for each inter-arrival and service time as follows. 
First, we set a time list $\mathbf{t}=[t_0,\,t_2,\,\cdots,\,t_n]$. 
The minimum time $t_0$ and the maximum time $t_n$ were set to 0 and the 99th percentile of the log-normal distribution, respectively.
The number of divisions $n+1$ was set to be sufficiently large, \textit{i.e.}, $\lfloor 12.5t_n+1\rfloor$, which represents the least integer greater than or equal to $12.5t_n+1$. 
Second, the cumulative density was configured as a list $\mathbf{\Phi}=[F(t_1),\,F(t_2),\,\cdots,\,F(t_n)]$, where $F(\cdot )$ is the cumulative density function of the log-normal distribution. 
Here, the probability density function and the cumulative density function of the log-normal distribution are represented as follows:
\begin{eqnarray}
f(x) &=& \frac{1}{\sqrt{2\pi}\sigma x} \exp\left(-\frac{(\ln x - \mu)^2}{2 \sigma^2}\right) \\
F(x) &=& \int_0^x f(x) \,\mathrm{d}x,
\end{eqnarray}
where $\mu$ and $\sigma$ are the mean and the standard deviation of the logarithm, respectively. 
Third, a uniform random number $r$ in the interval $[0,\,1)$ was introduced. 
Finally, we determined the time. 
If $r$ was in $[F(t_i),\,F(t_{i+1}))$, we defined the inter-arrival or the service time as $t_{i+1}$. 
If $r$ was greater than $F(t_n)$ the time was determined to be $t_n$. 

\vspace*{4pt}
\begin{figure}[htb]
	\begin{center}
		\includegraphics[width=0.7\hsize]{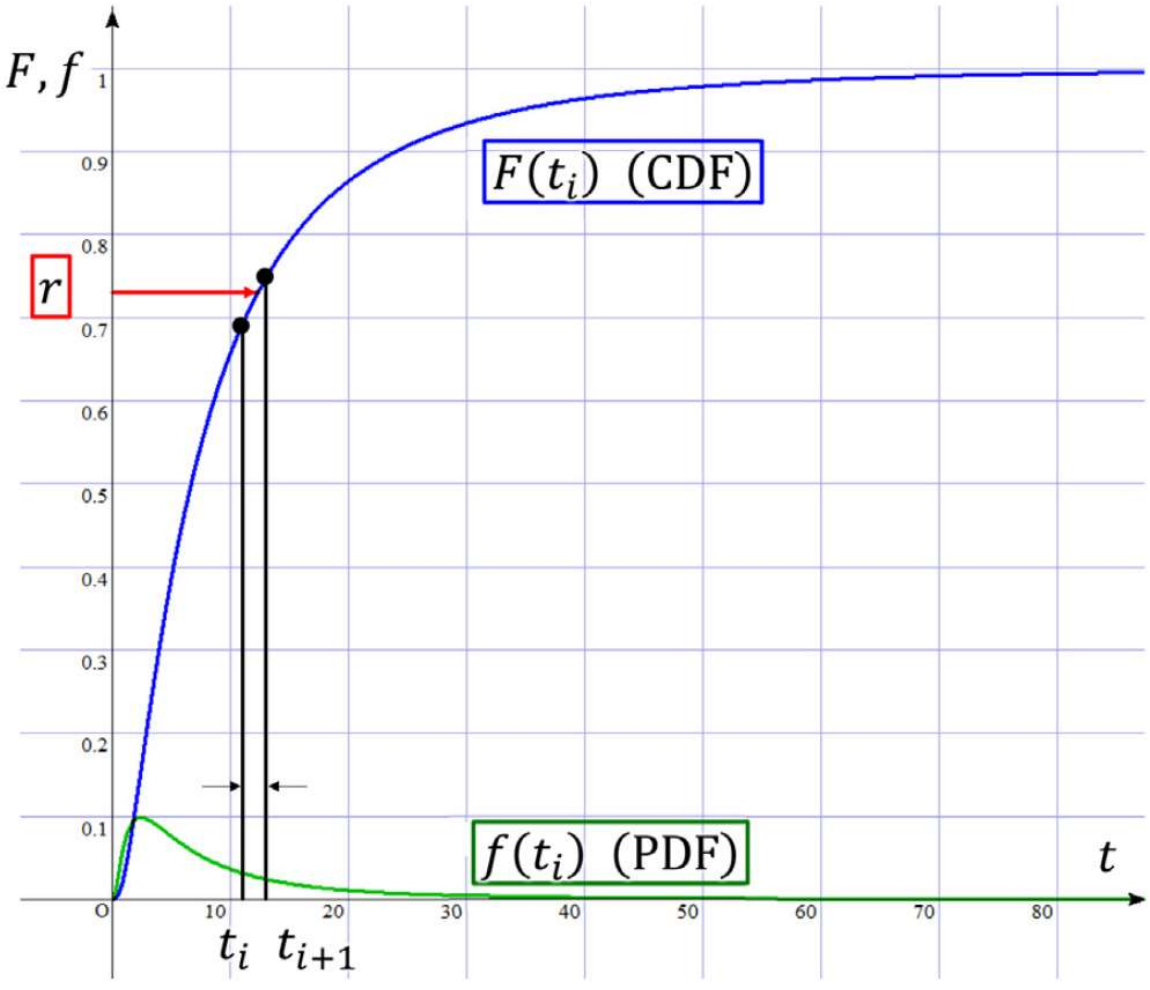}
	\end{center}
	\caption[Schematic of the calculation of the inter-arrival and service times from the log-normal distributions]{Schematic of the calculation of the inter-arrival and service times from the log-normal distributions in our simulation. The parameters of the log-normal distribution are set to $\mu=1.9$ and $\sigma=1.0$ in this figure.}
	\label{fig:Lognormal calculation method}
\end{figure}
\vspace*{4pt}

\end{document}